# Rapid synthesis of dual-element isotope-enriched α-MoO$_3$ crystals by reactive vapor transport

Ryan W. Spangler,[1] Jacob M. Shusterman,[2] Anton V. Ievlev,[2] Patrick E. Hopkins,[3,4,5] Joshua D. Caldwell,[6,7] Jon-Paul Maria[1]

1. Department of Materials Science and Engineering, The Pennsylvania State University, University Park, PA 16802, USA
2. Center for Nanophase Material Sciences, Oak Ridge National Laboratory, Oak Ridge, TN 37830, USA
3. Department of Mechanical and Aerospace Engineering, University of Virginia, Charlottesville, VA 22904, USA
4. Department of Materials Science and Engineering, University of Virginia, Charlottesville, VA 22904, USA
5. Department of Physics, University of Virginia, Charlottesville, VA 22904, USA
6. Interdisciplinary Materials Science Program, Vanderbilt University, Nashville, TN 37240, USA
7. Department of Mechanical Engineering, Vanderbilt University, Nashville, TN 37235, USA

Corresponding author: Jon-Paul Maria jpm133@psu.edu

## Abstract

In this work, we develop a rapid reactive vapor transport technique to efficiently utilize limited isotopically pure precursors, particularly gaseous $^{18}O_2$, and synthesize mm-scale, high-quality crystals within few-minute growth durations. We unlock this capability by using metallic molybdenum precursors with high source temperatures (900 °C) and total pressures (~1 atm) to maximize precursor efficiency and yield. Subsequently, we grow α-MoO$_3$ single crystals with high and uniform enrichment levels of $^{98}$Mo and $^{18}$O isotopes in several different permutations. As probed by Raman spectroscopy, modest and significant phonon energy redshifts occur following $^{98}$Mo and $^{18}$O enrichment, respectively. By demonstrating control over both molybdenum and oxygen isotopic fractions, we establish a powerful tool to advance nanophotonics and thermal management goals using α-MoO$_3$. This work is motivated by the possibility to enhance and engineer lattice vibrational mode phenomena including thermal conduction and hyperbolic phonon polariton (HPhP) dispersion—with particular interest in comparing the effects of light and heavy element enrichment.

## 1. Introduction

Isotope engineering can tune and optimize phonon-derived phenomena, including thermal conductivity and electron-phonon interactions, by altering elemental atomic masses and nuclear spins.[1–4] For instance, exchanging one isotope for another shifts phonon frequencies, while mixing multiple isotopic species increases phonon scattering events.[5–8] We can exploit these basic modifications to tailor more exotic phenomena as well. In certain strongly anisotropic single crystals with polar bonding, pairs of transverse optical (TO) and longitudinal optical (LO) phonon modes split, resulting in "hyperbolic" spectral regions wherein two principal components of the permittivity tensor have opposite signs.[9–11] These regions, termed hyperbolic Reststrahlen bands, host numerous exotic light-matter behaviors unique to hyperbolic media including low-loss, propagating hyperbolic phonon polariton modes (HPhPs)—infrared photons coupled with optical phonons.[12–18] Hyperbolicity was first realized using plasmon polaritons in nanostructured metallic-dielectric metamaterials;[19–21] recently, interest has expanded to include layered van der Waals (vdW) materials like hexagonal boron nitride (h-BN) and orthorhombic molybdenum trioxide (α-MoO$_3$), which naturally exhibit hyperbolic Reststrahlen bands without nanostructuring demands.[16–18,22,23] Compared to h-BN, α-MoO$_3$, a biaxially anisotropic material, commands additional in-plane directional control over polariton propagation and could advance mid-infrared (mid-IR) nanophotonics technologies including chemical sensing, planar focusing, and thermal management.[24–29] However, whereas the flexibility and spectrum of metamaterial design provides high tunability,[30,31] alternative frameworks are required to tailor naturally hyperbolic materials. For phonon-derived hyperbolic media like h-BN and α-MoO$_3$, isotope engineering is uniquely well suited for tuning HPhP modes to individual purposes.

Recent research, primarily using h-BN, has potently tailored HPhP modes through isotope engineering. Boron is naturally abundant as ~80% $^{11}$B and ~20% $^{10}$B; by synthesizing isotopically pure h-$^{11}$BN and h-$^{10}$BN, the HPhP lifetime can be improved manyfold by reducing phonon scattering losses.[32–34] Additionally, the Reststrahlen band frequencies can be considerably red-shifted by increasing the $^{10}$B/$^{11}$B ratio,[32,35] enabling HPhP dispersion engineering and even hetero-isotope nanostructure design.[36] The naturally abundant $^{14}$N can also be replaced with $^{15}$N (0.4% abundance) to further control HPhP modes.[35,37] In the case of α-MoO$_3$, molybdenum exists in seven abundant isotopes ranging from $^{92}$Mo to $^{100}$Mo, which introduce significant mass differences that increase phonon scattering losses. Previous studies on isotopically pure α-$^{92}$MoO$_3$ and α-$^{100}$MoO$_3$ indicate improved HPhP lifetimes, an over two-fold increase in propagation length, and modest Reststrahlen band shifts compared to naturally abundant crystals.[38,39] The oxygen isotope fraction, however, is an unexplored avenue for α-MoO$_3$; compared to $^{16}$O, which makes up 99.8% of naturally occurring oxygen, $^{18}$O is accompanied by a 12.5% mass increase. Due to the strong mass dependence of optical phonon frequencies,[40,41] replacing $^{16}$O with the heavier $^{18}$O could significantly alter the α-MoO$_3$ phonon modes and associated Reststrahlen bands. First, however, a process must be developed to synthesize α-MoO$_3$ crystals with control over both molybdenum and oxygen isotopic enrichments.

Large-area α-MoO$_3$ single crystals are readily synthesized using physical vapor transport techniques wherein a MoO$_3$ powder source, which has a high vapor pressure above ~700 °C,[42,43] is sublimated and subsequently deposited within a temperature gradient.[44–48] This technique can produce mm- to cm-scale single-crystal flakes which are several micrometers in thickness and are easily exfoliated. The molybdenum isotope fraction can be controlled by using commercially available molybdenum-isotope-enriched MoO$_3$ as the evaporation source; however, the reported processes require at least 8 hr for crystal growth.[38,39] Moreover, $^{18}$O enrichment has not yet been demonstrated due to the unavailability of $^{18}$O-enriched MoO$_3$ sources and the prohibitive expense of $^{18}$O$_2$ gas cylinders and the practical challenges of working with very small quantities in an atmospheric-pressure process. In this work, we use reactive sublimation to synthesize α-MoO$_3$ single crystals with controlled molybdenum and oxygen isotopic fractions by using a metallic molybdenum source, which has a much lower vapor pressure than α-MoO$_3$.[49–51] We then rapidly commence the evaporation-deposition process by oxidizing the molybdenum source with O$_2$ flow of select isotopic enrichment. Our method efficiently utilizes precursors, enabling high-yield crystal growth using less than 1 standard L of O$_2$ and 40 mg of molybdenum. We first develop the reactive vapor transport method using naturally abundant molybdenum and oxygen precursors and confirm the α-MoO$_3$ high crystallinity and proper stoichiometry using X-ray diffraction (XRD), Raman spectroscopy, energy-dispersive X-ray spectroscopy (EDS), and atomic force microscopy (AFM). We also control the α-MoO$_3$ flake size and morphology by modifying the total deposition pressure. We then synthesize single- and dual-element isotopically enriched α-MoO$_3$ flakes using pure $^{98}$Mo and $^{18}$O$_2$ sources and confirm the high attained isotopic enrichments using time-of-flight secondary ion mass spectrometry (ToF-SIMS), unlocking practical $^{18}$O-isotope engineering in α-MoO$_3$. Finally, using Raman spectroscopy, we investigate the effects of different isotopes on the vibrational properties of α-MoO$_3$ with implications for Reststrahlen band and HPhP engineering.

## 2. Methods

### 2.1 Reactive vapor transport synthesis

We perform the reactive vapor transport growths in a three-zone horizontal tube furnace (Thermcraft XST-3-0-24-3V2-F02) shown schematically in Fig. 1(a), where a molybdenum source is reactively sublimated in a hot zone (Zone 1) and transported to a cooler region (Zones 2-3) for deposition. The furnace is equipped with a 14 mm outer diameter (OD) quartz tube positioned coaxially within a 50 mm OD outer tube; the smaller tube serves as the growth chamber to maximize vapor concentrations *via* volume confinement to increase reaction and growth efficiencies. A rotary vane pump and mass flow controllers (MFCs) for Ar and $O_2$ gas flow are used to manage pressure and enable the controlled oxidation of the molybdenum metal source; we note that our reported pressures are approximate due to the dependence of our Convectron pressure gauge on gas composition. Prior to growth, 40 mg of Mo foil (Thermo Scientific, 99.95% purity) or $^{98}$Mo powder (IsoFlex, 99.998% purity, 98.65% enrichment) in a platinum boat is positioned in the smaller tube at Zone 1. For $^{18}$O-enriched growths, the chamber and gas lines are evacuated overnight to 12 mTorr to remove $^{16}O_2$ and adsorbed water. We show a typical experimental temporal temperature profile in Fig. 1(b) and summarize the different growth stages in Table I. During the heating stage (Stage I), we sustain a low-pressure argon environment to protect the source molybdenum from oxidizing environments. Immediately before deposition, we increase the argon flow to 500 sccm and meter the vacuum pumping speed to increase the total pressure to nearly 1 atm. In Stage II, to oxidize the molybdenum source and initiate $MoO_3$ evaporation and transport, we supplant the argon flow with 500 sccm of $^{16}O_2$ or $^{18}O_2$ (Millipore Sigma, 99% purity, 97-99% enrichment) for 2 min. For growth using $^{18}O_2$, due to the limited quantity and low pressure (2.3 atm) of the oxygen supply, the flow rate will decrease after approximately 75 s; at this point, we supplement the gas flow with 500 sccm of argon to transport the remaining $MoO_3$ vapor (from the rapidly oxidized Mo metal) to the deposition zone. In Stage III, we continue argon flow and cool the furnace naturally before venting to atmosphere, whereupon numerous large flakes are found on the interior walls of the inner quartz tube. We report four α-$MoO_3$ isotopic permutations: naturally abundant ($^{nat}Mo^{nat}O_3$), $^{98}$Mo-enriched ($^{98}Mo^{nat}O_3$), $^{18}$O-enriched ($^{nat}Mo^{18}O_3$), and dual-element isotope-enriched ($^{98}Mo^{18}O_3$). For the $^{98}Mo^{nat}O_3$ sample, 50 mg of $^{98}$Mo-enriched $MoO_3$ (IsoFlex, 99.9% purity, 98.42% enrichment) is the source powder. Since no reactive sublimation is necessary for this oxide source, we heat Zone 1 to 800 °C and flow 50 sccm $N_2$ flow at atmospheric pressure for a 50 min deposition. After each growth, the furnace zones are heated to high temperatures (950 °C) to remove volatile deposition products.

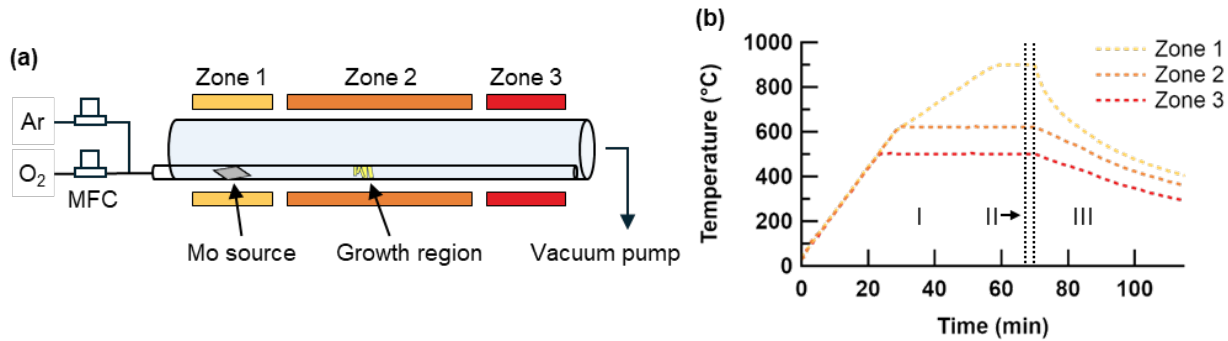

**Figure 1. Reactive vapor transport growth method.** (a) Schematic of the growth chamber used for crystal growth. (b) Temperature profile of the three-zone furnace during a typical growth run comprised of the ramp-up (I), growth (II), and cool-down (III) stages.

**Table I.** Parameters for each heating stage in the reactive vapor transport process

| Stage | Duration (min) | Ar (sccm) | $O_2$ (sccm) | Pressure (Torr) |
|---|---|---|---|---|
| I – Ramp up | 65 | 50 | 0 | 0.07 |
| II – Growth | 2 | 0 | 500 | 150-740 |
| III – Cool down | 120 | 500 | 0 | 740 |

### 2.2 Characterization

An Olympus BX60M bright-field optical microscope is used to visually inspect the α-MoO₃ crystals. Energy-dispersive X-ray spectrometry (EDS) is performed using a Tescan Mira scanning electron microscope (SEM) and EDS detector with an accelerating voltage of 15 keV and beam current of 1 nA. Our reported Mo:O ratio is averaged from five EDS point measurements with carbon excluded. The surface morphology is characterized by using an Asylum MFP-3D atomic force microscope (AFM) in tapping mode. X-ray diffraction is performed using a Panalytical Empyrean diffractometer with a Cu source and Bragg-Brentano[HD] incident beam optics. Raman spectroscopy is performed using a Horiba LabRAM HR Evolution microscope without a polarization analyzer equipped with a 532 nm excitation laser and an 1800 gr/mm grating. To account for the anisotropic Raman response of α-MoO₃,[52,53] the crystal *a*-axis is oriented parallel to the linear polarization of the incident laser.

Time-of-flight secondary ion mass spectrometry (ToF-SIMS) measurements were performed using a TOF.SIMS⁵-NSC instrument (ION-TOF GmbH, Germany) operated in positive and negative ion detection modes to measure the molybdenum and oxygen isotopic fractions, respectively. The $Bi_3^+$ ion beam was used as a primary ion source and operated at 30 keV energy and 0.5 nA current in DC mode with a focused ion beam spot size of approximately 120 nm. Imaging was performed over a 100 x 100 μm area of 256 x 256 pixels with acquisition time of 19.16 s. Depth profiling was performed with a $Cs^+$ ion-sputtering source for both positive and negative ion detection modes operated at 500 eV energy with DC currents ranging from 35 - 45 nA. The ion-sputtering source was rastered across a 500 x 500 μm area with 10 sputter frames per analysis scan. A low energy electron flood gun was used during SIMS measurements to compensate for surface charging effects of the primary ion gun.

## 3. Results and discussion

### 3.1 Growth using naturally abundant isotopes

We first use naturally abundant molybdenum and $O_2$ precursors to develop the high-efficiency reactive vapor deposition process. We visually monitor the growth process *in situ* through a viewport on one end of the tube furnace. The high source temperature of 900 °C maximizes growth rates by vaporizing the molybdenum as quickly as it oxidizes: approximately 10 s after commencing $O_2$ flow, the metal source begins oxidizing and glows a bright yellow-orange color for 10 s before vanishing as the newly formed $MoO_3$ fully vaporizes into a whitish cloud. Soon afterwards, small crystals first become visible on the smaller tube interior and continue to grow for several minutes as the remaining $MoO_3$ vapor is transported to the deposition zone, with growth briefly continuing during cooling in flowing Ar. Afterwards, approximately 10-20 large flakes (>3 mm in at least one lateral dimension) and dozens of smaller flakes (>100 μm) can be harvested from the smaller tube interior as shown in Fig. 2(a).

The flake morphologies range from needles to plates and are typical of α-$MoO_3$ grown using PVT techniques, with frequent faceting along the *c*- and *a*-axes as in the optical micrograph in Fig. 2(b).[48] The flake thicknesses vary considerably from several hundreds of nanometers to a few micrometers; regardless, isolating, transferring, and exfoliating thin, smooth samples is facile. We confirmed the α-$MoO_3$ stoichiometry in Fig. 2(c), which shows a representative EDS spectrum with an approximately 3:1 O:Mo ratio.

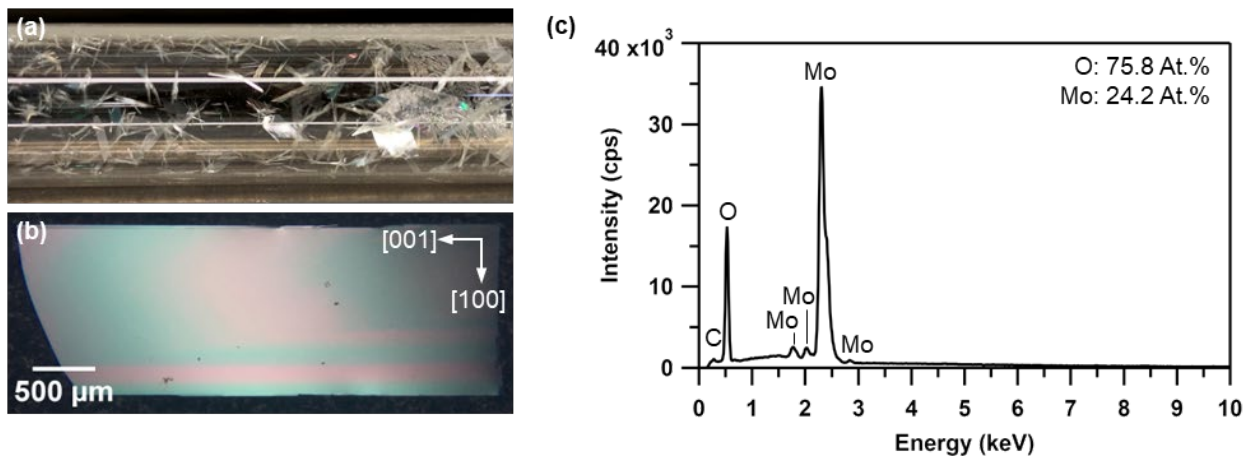

**Figure 2. α-MoO₃ flakes.** (a) Photograph of numerous large crystals grown within the small quartz tube. (b) Optical micrograph and (c) EDS spectrum and stoichiometry of the α-MoO₃ crystals.

The α-$MoO_3$ crystal structure, sourced from the Materials Project database (ID: mp-20589),[54,55] is shown in Fig. 3(a) and is described by an orthorhombic unit cell (space group *Pbnm*) with the lattice parameters $a$ = 3.962 Å, $b$ = 13.860 Å, and $c$ = 3.697 Å.[56,57] The unit cell is composed of distorted Mo-$O_6$ octahedra which are strongly bonded in the *a-c* plane and ordered into vdW bilayers along the *b*-axis. In Fig. 3(b), we show a representative XRD scan of a flake transferred onto a sapphire substrate. Only sharp, intense α-MoO$_3$ (0*k*0) reflections are observed, indicating phase purity and high crystallinity and confirming that the *b*-axis is oriented parallel to the flake thickness. The Raman spectrum shown in Fig. 3(c) displays narrow, intense peaks that are also characteristic of α-MoO$_3$,[52,58] further indicating the high structural quality. Dieterle *et al*.[58] found that the intensity ratio of the Raman bands at 295 cm$^{-1}$ and 285 cm$^{-1}$ can be used to indicate oxygen sub-stoichiometry, which can be common in α-MoO$_3$.[59] We do not observe the 295 cm$^{-1}$ peak in our results, agreeing with the EDS analysis in Fig. 3(c) that our α-MoO$_3$ are close to stoichiometric. The AFM image shown in Fig. 3(d) shows broad terraces with clean, linear steps. The step height is approximately 0.70-0.75 nm, in agreement or slightly larger than the expected *b*/2 interlayer spacing of 0.693 nm.[56,57] Here, the terrace widths alternate between wide (~3.4 µm) and narrow (~0.35 µm), although the absolute widths vary at different positions along the crystal. The alternating terrace widths may indicate different step migration or re-evaporation rates of the (010) and (020) planes, although further research is necessary to fully examine the surface dynamics during growth.

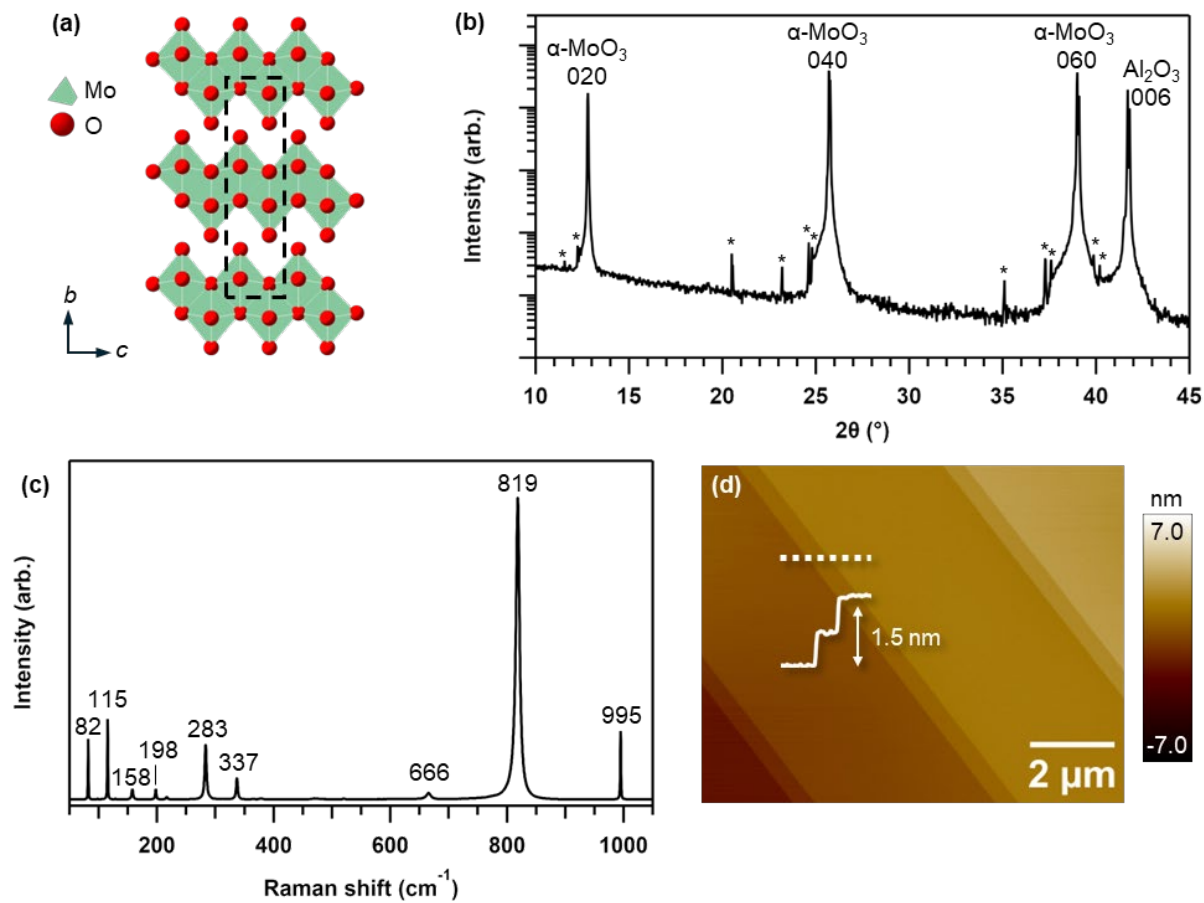

**Figure 3. Physical properties of α-MoO₃ flakes.** (a) α-MoO₃ crystal structure viewed along the a-axis with the unit cell outlined. (b) XRD θ-2θ scan of a α-MoO₃ flake transferred onto a c-axis-oriented Al₂O₃ substrate. (*) = reflections arising from secondary and $\frac{\lambda}{2}$ wavelengths. (c) Raman spectrum of a flake transferred onto a Si substrate with the major α-MoO₃ vibration frequencies labeled, and (d) AFM height image and (inset) line profile.

We also studied the total growth pressure influence on the flake morphology by growing in oxygen environments at 150, 360, 490, and 740 Torr, where the resulting flake yields are photographed in Fig. 4(a-d), respectively. At 150 Torr, the reaction rate is low, causing only a minor (~10 mg) molybdenum source mass loss and a relatively low yield of small (<500 μm) crystals. As we increase the growth pressure, the molybdenum source is fully consumed and the flake size increases significantly, with the largest crystals grown at 740 Torr. Additionally, the growth region shifts to the left by ~3 cm as the pressure is increased from 150–740 Torr as nucleation and growth stabilize at higher temperatures. Despite conventional wisdom assuring that high precursor supersaturations typically favor nucleation over crystal growth,[60] these results agree with other reported α-MoO₃ vapor transport growths where high vapor concentrations promote larger flakes.[48] The various complexities of the growth process may collectively trigger this counterintuitive relation. In particular, the axial flow velocity decreases at higher total pressures, leading to longer residence times which increases the duration of the growth stage. Notably, we directly observe this effect by visually monitoring the crystal growth onset and termination times, which separate and broaden to allow prolonged growth at higher total pressures.

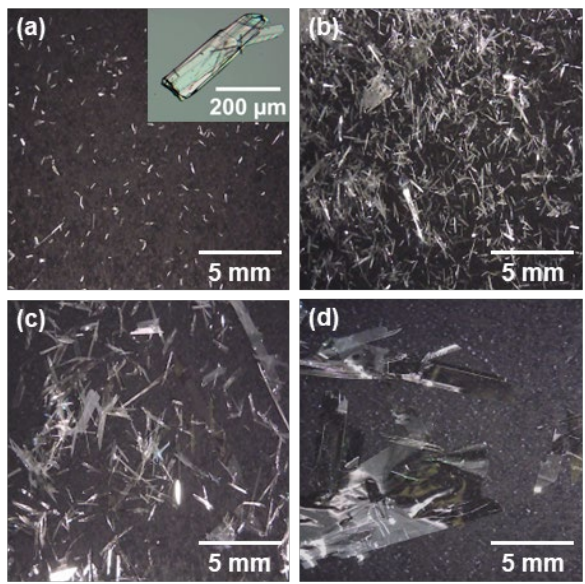

**Figure 4. Pressure dependence of reactive vapor transport growth.** (a-d) Photographs and (inset) optical micrograph of α-MoO$_3$ flakes grown at different O$_2$ pressures: (a) 150 Torr, (b) 360 Torr, (c) 490 Torr, and (d) 740 Torr.

### 3.2 Growth using enriched $^{98}$Mo and $^{18}$O precursors

With a sound understanding of and control over the process for large-area, high-quality α-MoO$_3$ crystal growth from a metal source, we can implement isotope-enriched $^{98}$Mo and $^{18}$O$_2$ precursors to engineer different isotopic combinations. Using the same methods for naturally abundant sample ($^{nat}$Mo$^{nat}$O$_3$) preparation, we produced $^{98}$Mo-enriched ($^{98}$Mo$^{nat}$O$_3$), $^{18}$O-enriched ($^{nat}$Mo$^{18}$O$_3$), and dual-element isotope-enriched ($^{98}$Mo$^{18}$O$_3$) flakes. The isotopic fractions of metal and oxygen are quantified by ToF-SIMS depth profiles. In Fig. 5(a), we show the $^{98}$Mo content of the molybdenum-enriched and unenriched samples as a function of sputter time. We observe a high, uniform, and identical $^{98}$Mo fraction of 98.7% for both $^{98}$MoO$_3$ and $^{98}$Mo$^{18}$O$_3$, while the $^{nat}$Mo$^{18}$O$_3$ matches the expected ~25% naturally abundant fraction of $^{98}$Mo. The oxygen-isotope measurements are more challenging to interpret as we observe occasional quantitative disagreement in measurements taken from different locations. To improve our confidence in the quantifications obtained, we measured three areas on different flakes of each isotopic permutation. In some of these scans, unexpected features including high C content near the top surface and odd $^{16/18}$O profile shapes cause uncertainty in the measurements. To investigate these inconsistencies, we performed Raman spectroscopy on numerous flakes and observed identical spectra—without peak shifting or broadening—across all measured flakes of each batch, confirming that the flakes are uniform and alike. Therefore, we believe that artifacts resulting from variable flake topography and tilt, for instance, result in artificially high $^{16}$O signals in some ToF-SIMS measurements.[61,62] In Fig. 5(b), we show the most self-consistent depth profiles of the $^{18}$O fractions taken from each sample batch. The near-identical shape and magnitude of the two $^{18}$O-enriched profiles give us confidence that the true isotopic fraction is likely ~87% $^{18}$O. We note that of the six measurements, the lowest enrichment level is ~64% $^{18}$O; therefore, even the more unreliable profiles still indicate a significant $^{18}$O enrichment compared to the 0.2% natural abundance. The reduced $^{18}$O-enrichment of the flakes compared to the 97-99% isotopically pure $^{18}$O$_2$ precursors could result from $^{16}$O present in the molybdenum precursor native oxide and/or residual water and oxygen in the growth chamber or gas lines. Regardless, the relatively high obtained enrichment levels illuminate the effectiveness of our reactive vapor transport technique to control both the metal- and oxygen-isotope ratios in large-area α-MoO$_3$ crystals.

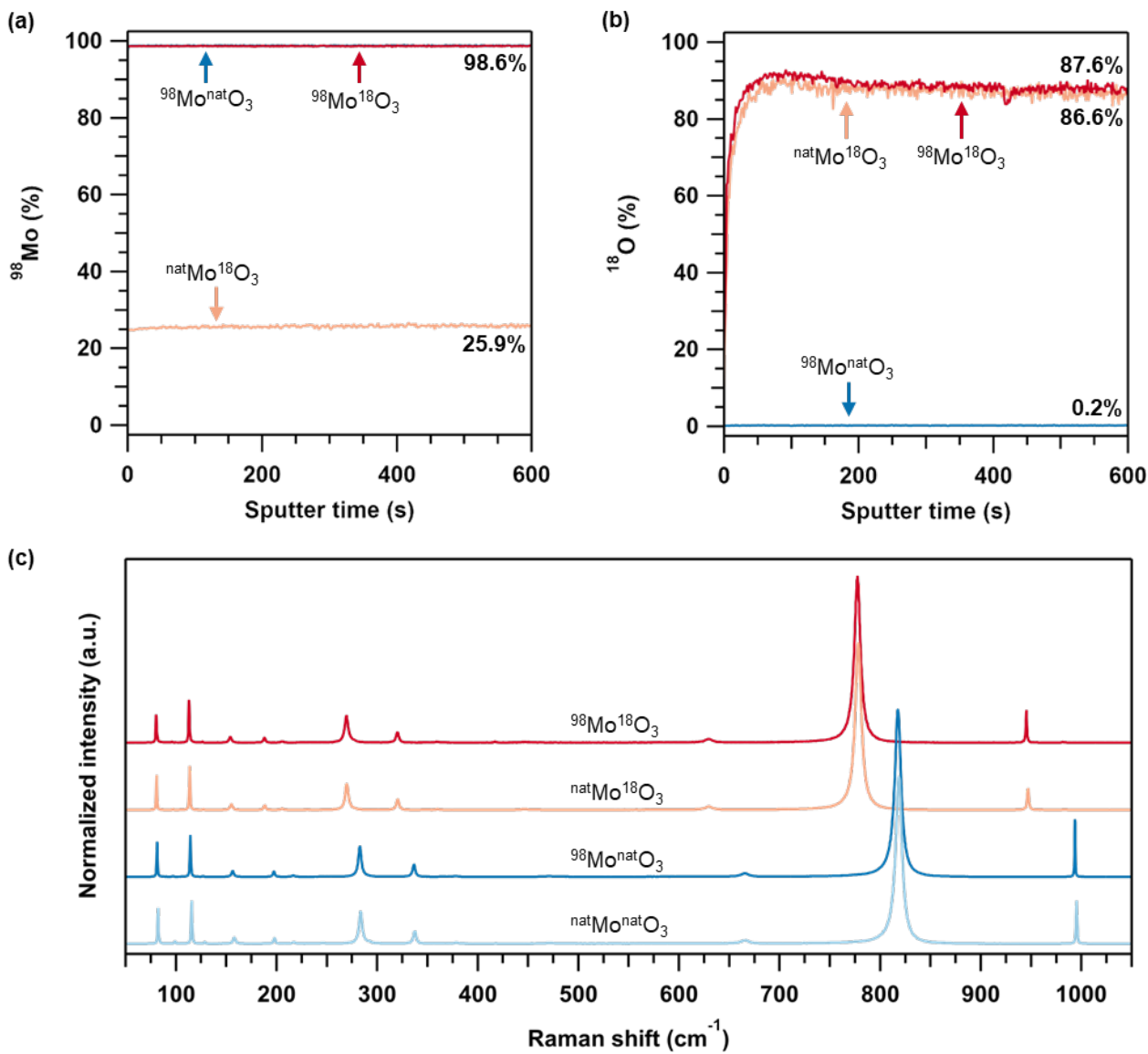

**Figure 5. Dual-element isotope enrichment of α-MoO$_3$.** Depth profiles of (a) $^{98}$Mo and (b) $^{18}$O isotope fractions collected by ToF-SIMS. In (a), the $^{98}$Mo$^{nat}$O$_3$ and $^{98}$Mo$^{18}$O$_3$ profiles overlap. (c) Offset Raman spectra of the naturally abundant and isotope-enriched α-MoO$_3$ flakes after normalization to the main α-MoO$_3$ peak (~820 cm$^{-1}$).

Fig. 5(c) shows the Raman spectra of the different isotopic α-MoO$_3$ permutations to elucidate how isotope enrichment modifies the phonon properties. The A$_g$ mode at 995 cm$^{-1}$ intensifies for $^{98}$Mo-enriched crystals compared to naturally abundant α-MoO$_3$, potentially due to reduced isotope scattering. All Raman peaks redshift modestly; for instance, A$_g$ 995 cm$^{-1}$ shifting by less than 2 cm$^{-1}$. This shift is smaller than reported $^{92}$Mo and $^{100}$Mo-enriched α-MoO$_3$, where larger changes in average atomic mass induce 3-4 cm$^{-1}$ shifts in either direction.[38] In contrast, $^{18}$O-enrichment induces a more significant spectral redshift: A$_g$ 995 cm$^{-1}$ moves to 947 cm$^{-1}$, a shift of nearly 50 cm$^{-1}$. The intensity of this peak reduces, most likely due to increased phonon scattering by the now co-existing $^{16}$O and $^{18}$O species. Effects compound in the dual-element isotope-enriched sample, resulting in a 945 cm$^{-1}$ high-energy A$_g$ peak with moderate intensity. Although the Reststrahlen bands cannot be extracted directly from Raman spectra of α-MoO₃, our results are consistent with the expected universal softening of phonon modes which applies to the Raman-inactive TO and LO phonon modes that bound the Reststrahlen bands.[32,38] Therefore, incorporating $^{18}$O induces stronger α-MoO₃ Reststrahlen band tunability than molybdenum isotopes can, where precisely modifying isotope fractions could realize broad spectral control of HPhP modes.

4. **Conclusions**

Using reactive vapor transport, we have developed a method to rapidly synthesize α-MoO₃ flakes with control over both the oxygen and molybdenum isotopic fractions. Our technique efficiently consumes limited $^{18}$O₂ sources to enable high-yield growth using less than 1 standard L of gaseous oxygen. Chemical and structural characterization reveals the high flake crystalline and stoichiometric quality. We vary the flake size and morphology by modifying the growth pressure, with synthesis at near-atmospheric-pressure producing large crystals comparable with those reported by other growth methods. ToF-SIMS analyses quantify the obtained isotopic fractions as 98.7% $^{98}$Mo and at least ~75% $^{18}$O. We find that $^{18}$O enrichment facilitates significant phonon mode tunability that exceeds the capabilities of molybdenum-isotope engineering. To the best of our knowledge, this is the first reported synthesis of highly $^{18}$O-enriched α-MoO₃—a capability critical to exploring cutting-edge phonon-derived phenomena in α-MoO₃ including enhanced thermal conductivity and tunable HPhP modes.

5. **Acknowledgements**

This work was supported by the Army Research Office under grant no. W911NF-21-1-0119. ToF-SIMS characterization was conducted at the Center for Nanophase Materials Sciences, which is a DOE Office of Science User Facility, and using instrumentation within ORNL's Materials Characterization Core provided by UT-Battelle, LLC under Contract No. DE-AC05-00OR22725 with the U.S. Department of Energy. PEH appreciates support from the Office of Naval Research, Grant No. N00014-26-1-2076. The authors also acknowledge help from the Penn State Materials Research Institute, particularly Maxwell Wetherington at the Penn State Materials Characterization Laboratory for helpful discussions on Raman spectroscopy.

# 6. References

1. Plekhanov, V. G. (2000). Isotope engineering. *Physics-Uspekhi*, *43*(11), 1147. https://doi.org/10.1070/PU2000v043n11ABEH000264
2. Chen, K., Song, B., Ravichandran, N. K., Zheng, Q., Chen, X., Lee, H., Sun, H., Li, S., Udalamatta Gamage, G. A. G., Tian, F., Ding, Z., Song, Q., Rai, A., Wu, H., Koirala, P., Schmidt, A. J., Watanabe, K., Lv, B., Ren, Z., … Chen, G. (2020). Ultrahigh thermal conductivity in isotope-enriched cubic boron nitride. *Science*, *367*(6477), 555–559. https://doi.org/10.1126/science.aaz6149
3. Vuong, T. Q. P., Liu, S., Van der Lee, A., Cuscó, R., Artús, L., Michel, T., Valvin, P., Edgar, J. H., Cassabois, G., & Gil, B. (2018). Isotope engineering of van der Waals interactions in hexagonal boron nitride. *Nature Materials*, *17*(2), 152–158. https://doi.org/10.1038/nmat5048
4. Yu, Y., Turkowski, V., Hachtel, J. A., Puretzky, A. A., Ievlev, A. V., Din, N. U., Harris, S. B., Iyer, V., Rouleau, C. M., Rahman, T. S., Geohegan, D. B., & Xiao, K. (2024). Anomalous isotope effect on the optical bandgap in a monolayer transition metal dichalcogenide semiconductor. *Science Advances*, *10*(8), eadj0758. https://doi.org/10.1126/sciadv.adj0758
5. Srivastava, G. P. (1990). *The Physics of Phonons* (1st ed.). Routledge. https://doi.org/10.1201/9780203736241
6. Abeles, B. (1963). Lattice Thermal Conductivity of Disordered Semiconductor Alloys at High Temperatures. *Physical Review*, *131*(5), 1906–1911. https://doi.org/10.1103/PhysRev.131.1906
7. Cahill, D. G., & Watanabe, F. (2004). Thermal conductivity of isotopically pure and Ge-doped Si epitaxial layers from 300 to 550 K. *Physical Review B*, *70*(23), 235322. https://doi.org/10.1103/PhysRevB.70.235322
8. Giri, A., & Hopkins, P. E. (2020). Achieving a better heat conductor. *Nature Materials*, *19*(5), 482–484. https://doi.org/10.1038/s41563-020-0658-z
9. Smith, D. R., & Schurig, D. (2003). Electromagnetic Wave Propagation in Media with Indefinite Permittivity and Permeability Tensors. *Physical Review Letters*, *90*(7), 077405. https://doi.org/10.1103/PhysRevLett.90.077405
10. Smith, D. R., Kolinko, P., & Schurig, D. (2004). Negative refraction in indefinite media. *Journal of the Optical Society of America B*, *21*(5), 1032. https://doi.org/10.1364/JOSAB.21.001032
11. Hoffman, A. J., Alekseyev, L., Howard, S. S., Franz, K. J., Wasserman, D., Podolskiy, V. A., Narimanov, E. E., Sivco, D. L., & Gmachl, C. (2007). Negative refraction in semiconductor metamaterials. *Nature Materials*, *6*(12), Article 12. https://doi.org/10.1038/nmat2033
12. Jacob, Z., Smolyaninov, I. I., & Narimanov, E. E. (2012). Broadband Purcell effect: Radiative decay engineering with metamaterials. *Applied Physics Letters*, *100*(18), 181105. https://doi.org/10.1063/1.4710548
13. Biehs, S.-A., Tschikin, M., & Ben-Abdallah, P. (2012). Hyperbolic Metamaterials as an Analog of a Blackbody in the Near Field. *Physical Review Letters*, *109*(10), Article 10. https://doi.org/10.1103/PhysRevLett.109.104301

14. Folland, T. G., Fali, A., White, S. T., Matson, J. R., Liu, S., Aghamiri, N. A., Edgar, J. H., Haglund, R. F., Abate, Y., & Caldwell, J. D. (2018). Reconfigurable infrared hyperbolic metasurfaces using phase change materials. *Nature Communications*, *9*(1), 4371. https://doi.org/10.1038/s41467-018-06858-y
15. He, M., Iyer, G. R. S., Aarav, S., Sunku, S. S., Giles, A. J., Folland, T. G., Sharac, N., Sun, X., Matson, J., Liu, S., Edgar, J. H., Fleischer, J. W., Basov, D. N., & Caldwell, J. D. (2021). Ultrahigh-Resolution, Label-Free Hyperlens Imaging in the Mid-IR. *Nano Letters*, *21*(19), 7921–7928. https://doi.org/10.1021/acs.nanolett.1c01808
16. Caldwell, J. D., Aharonovich, I., Cassabois, G., Edgar, J. H., Gil, B., & Basov, D. N. (2019). Photonics with hexagonal boron nitride. *Nature Reviews Materials*, *4*(8), 552–567. https://doi.org/10.1038/s41578-019-0124-1
17. Dai, S., Fei, Z., Ma, Q., Rodin, A. S., Wagner, M., McLeod, A. S., Liu, M. K., Gannett, W., Regan, W., Watanabe, K., Taniguchi, T., Thiemens, M., Dominguez, G., Neto, A. H. C., Zettl, A., Keilmann, F., Jarillo-Herrero, P., Fogler, M. M., & Basov, D. N. (2014). Tunable Phonon Polaritons in Atomically Thin van der Waals Crystals of Boron Nitride. *Science*, *343*(6175), Article 6175. https://doi.org/10.1126/science.1246833
18. Caldwell, J. D., Kretinin, A. V., Chen, Y., Giannini, V., Fogler, M. M., Francescato, Y., Ellis, C. T., Tischler, J. G., Woods, C. R., Giles, A. J., Hong, M., Watanabe, K., Taniguchi, T., Maier, S. A., & Novoselov, K. S. (2014). Sub-diffractional volume-confined polaritons in the natural hyperbolic material hexagonal boron nitride. *Nature Communications*, *5*(1), 5221. https://doi.org/10.1038/ncomms6221
19. Cortes, C. L., Newman, W., Molesky, S., & Jacob, Z. (2012). Quantum nanophotonics using hyperbolic metamaterials. *Journal of Optics*, *14*(6), Article 6. https://doi.org/10.1088/2040-8978/14/6/063001
20. Poddubny, A., Iorsh, I., Belov, P., & Kivshar, Y. (2013). Hyperbolic metamaterials. *Nature Photonics*, *7*(12), 948–957. https://doi.org/10.1038/nphoton.2013.243
21. Kildishev, A. V., Boltasseva, A., & Shalaev, V. M. (2013). Planar Photonics with Metasurfaces. *Science*, *339*(6125), Article 6125. https://doi.org/10.1126/science.1232009
22. Ma, W., Alonso-González, P., Li, S., Nikitin, A. Y., Yuan, J., Martín-Sánchez, J., Taboada-Gutiérrez, J., Amenabar, I., Li, P., Vélez, S., Tollan, C., Dai, Z., Zhang, Y., Sriram, S., Kalantar-Zadeh, K., Lee, S.-T., Hillenbrand, R., & Bao, Q. (2018). In-plane anisotropic and ultra-low-loss polaritons in a natural van der Waals crystal. *Nature*, *562*(7728), 557–562. https://doi.org/10.1038/s41586-018-0618-9
23. Zheng, Z., Chen, J., Wang, Y., Wang, X., Chen, X., Liu, P., Xu, J., Xie, W., Chen, H., Deng, S., & Xu, N. (2018). Highly Confined and Tunable Hyperbolic Phonon Polaritons in Van Der Waals Semiconducting Transition Metal Oxides. *Advanced Materials*, *30*(13), 1705318. https://doi.org/10.1002/adma.201705318
24. Álvarez-Pérez, G., Folland, T. G., Errea, I., Taboada-Gutiérrez, J., Duan, J., Martín-Sánchez, J., Tresguerres-Mata, A. I. F., Matson, J. R., Bylinkin, A., He, M., Ma, W., Bao, Q., Martín, J. I., Caldwell, J. D., Nikitin, A. Y., & Alonso-González, P. (2020). Infrared Permittivity of the Biaxial van der Waals Semiconductor α-MoO3 from Near- and Far-Field Correlative Studies. *Advanced Materials*, *32*(29), Article 29. https://doi.org/10.1002/adma.201908176
25. Zheng, Z., Xu, N., Oscurato, S. L., Tamagnone, M., Sun, F., Jiang, Y., Ke, Y., Chen, J., Huang, W., Wilson, W. L., Ambrosio, A., Deng, S., & Chen, H. (2019). A mid-infrared biaxial hyperbolic van der Waals crystal. *Science Advances*, *5*(5), eaav8690. https://doi.org/10.1126/sciadv.aav8690
26. Barcelos, I. D., Canassa, T. A., Mayer, R. A., Feres, F. H., de Oliveira, E. G., Goncalves, A.-M. B., Bechtel, H. A., Freitas, R. O., Maia, F. C. B., & Alves, D. C. B. (2021). Ultrabroadband Nanocavity of Hyperbolic Phonon–Polaritons in 1D-Like α-MoO3. *ACS Photonics*, *8*(10), 3017–3026. https://doi.org/10.1021/acsphotonics.1c00955
27. Qu, Y., Chen, N., Teng, H., Hu, H., Sun, J., Yu, R., Hu, D., Xue, M., Li, C., Wu, B., Chen, J., Sun, Z., Liu, M., Liu, Y., García de Abajo, F. J., & Dai, Q. (2022). Tunable Planar Focusing Based on Hyperbolic Phonon Polaritons in α-MoO3. *Advanced Materials*, *34*(23), 2105590. https://doi.org/10.1002/adma.202105590
28. He, M., Folland, T. G., Duan, J., Alonso-González, P., De Liberato, S., Paarmann, A., & Caldwell, J. D. (2022). Anisotropy and Modal Hybridization in Infrared Nanophotonics Using Low-Symmetry Materials. *ACS Photonics*, *9*(4), 1078–1095. https://doi.org/10.1021/acsphotonics.1c01486
29. Chen, Y., Pacheco, M. A. S., Salihoglu, H., & Xu, X. (2024). Greatly Enhanced Radiative Transfer Enabled by Hyperbolic Phonon Polaritons in α-MoO3. *Advanced Functional Materials*, *34*(40), 2403719. https://doi.org/10.1002/adfm.202403719
30. Cortes, C. L., Newman, W., Molesky, S., & Jacob, Z. (2012). Quantum nanophotonics using hyperbolic metamaterials. *Journal of Optics*, *14*(6), 063001. https://doi.org/10.1088/2040-8978/14/6/063001
31. Cleri, A. J., Nolen, J. R., Wirth, K. G., He, M., Runnerstrom, E. L., Kelley, K. P., Nordlander, J., Taubner, T., Folland, T. G., Maria, J.-P., & Caldwell, J. D. (2023). Tunable, Homoepitaxial Hyperbolic Metamaterials Enabled by High Mobility CdO. *Advanced Optical Materials*, *11*(1), Article 1. https://doi.org/10.1002/adom.202202137
32. Giles, A. J., Dai, S., Vurgaftman, I., Hoffman, T., Liu, S., Lindsay, L., Ellis, C. T., Assefa, N., Chatzakis, I., Reinecke, T. L., Tischler, J. G., Fogler, M. M., Edgar, J. H., Basov, D. N., & Caldwell, J. D. (2018). Ultralow-loss polaritons in isotopically pure boron nitride. *Nature Materials*, *17*(2), Article 2. https://doi.org/10.1038/nmat5047
33. Pavlidis, G., Schwartz, J. J., Matson, J., Folland, T., Liu, S., Edgar, J. H., Caldwell, J. D., & Centrone, A. (2021). Experimental confirmation of long hyperbolic polariton lifetimes in monoisotopic (10B) hexagonal boron nitride at room temperature. *APL Materials*, *9*(9), 091109. https://doi.org/10.1063/5.0061941
34. Ni, G., McLeod, A. S., Sun, Z., Matson, J. R., Lo, C. F. B., Rhodes, D. A., Ruta, F. L., Moore, S. L., Vitalone, R. A., Cusco, R., Artús, L., Xiong, L., Dean, C. R., Hone, J. C., Millis, A. J., Fogler, M. M., Edgar, J. H., Caldwell, J. D., & Basov, D. N. (2021). Long-Lived Phonon Polaritons in Hyperbolic Materials. *Nano Letters*, *21*(13), 5767–5773. https://doi.org/10.1021/acs.nanolett.1c01562
35. He, M., Lindsay, L., Beechem, T. E., Folland, T., Matson, J., Watanabe, K., Zavalin, A., Ueda, A., Collins, Warren. E., Taniguchi, T., & Caldwell, J. D. (2021). Phonon engineering of boron nitride via isotopic enrichment. *Journal of Materials Research*, *36*(21), 4394–4403. https://doi.org/10.1557/s43578-021-00426-9
36. Chen, M., Zhong, Y., Harris, E., Li, J., Zheng, Z., Chen, H., Wu, J.-S., Jarillo-Herrero, P., Ma, Q., Edgar, J. H., Lin, X., & Dai, S. (2023). Van der Waals isotope heterostructures for engineering phonon polariton dispersions. *Nature Communications*, *14*(1), 4782. https://doi.org/10.1038/s41467-023-40449-w

37. Janzen, E., Schutte, H., Plo, J., Rousseau, A., Michel, T., Desrat, W., Valvin, P., Jacques, V., Cassabois, G., Gil, B., & Edgar, J. H. (2024). Boron and Nitrogen Isotope Effects on Hexagonal Boron Nitride Properties. *Advanced Materials*, *36*(2), 2306033. https://doi.org/10.1002/adma.202306033
38. Zhao, Y., Chen, J., Xue, M., Chen, R., Jia, S., Chen, J., Bao, L., Gao, H.-J., & Chen, J. (2022). Ultralow-Loss Phonon Polaritons in the Isotope-Enriched α-MoO$_3$. *Nano Letters*, *22*(24), 10208–10215. https://doi.org/10.1021/acs.nanolett.2c03742
39. Schultz, J. F., Krylyuk, S., Schwartz, J. J., Davydov, A. V., & Centrone, A. (2024). Isotopic effects on in-plane hyperbolic phonon polaritons in MoO3. *Nanophotonics*, *13*(9), 1581–1592. https://doi.org/10.1515/nanoph-2023-0717
40. Kittel, C., & Masi, J. F. (1954). Introduction to Solid State Physics. *Physics Today*, *7*(8), 18–19. https://doi.org/10.1063/1.3061720
41. Caldwell, J. D., Lindsay, L., Giannini, V., Vurgaftman, I., Reinecke, T. L., Maier, S. A., & Glembocki, O. J. (2015). Low-loss, infrared and terahertz nanophotonics using surface phonon polaritons. *Nanophotonics*, *4*(1), 44–68. https://doi.org/10.1515/nanoph-2014-0003
42. Berkowitz, J., Inghram, M. G., & Chupka, W. A. (1957). Polymeric Gaseous Species in the Sublimation of Molybdenum Trioxide. *The Journal of Chemical Physics*, *26*(4), 842–846. https://doi.org/10.1063/1.1743417
43. Blackburn, P. E., Hoch, M., & Johnston, H. L. (1958). The Vaporization of Molybdenum and Tungsten Oxides. *The Journal of Physical Chemistry*, *62*(7), 769–773. https://doi.org/10.1021/j150565a001
44. Fourcaudot, G., Gourmala, M., & Mercier, J. (1979). Vapor phase transport and crystal growth of molybdenum trioxide and molybdenum ditelluride. *Journal of Crystal Growth*, *46*(1), 132–135. https://doi.org/10.1016/0022-0248(79)90120-9
45. Zeng, H. C. (1998). Chemical Etching of Molybdenum Trioxide: A New Tailor-Made Synthesis of MoO3 Catalysts. *Inorganic Chemistry*, *37*(8), 1967–1973. https://doi.org/10.1021/ic971269v
46. Balakumar, S., & Zeng, H. C. (1999). Growth modes in vapour-phase prepared orthorhombic molybdenum trioxide crystals. *Journal of Crystal Growth*, *197*(1), 186–194. https://doi.org/10.1016/S0022-0248(98)00924-5
47. Zheng, Q., Huang, J., Cao, S., & Gao, H. (2015). A flexible ultraviolet photodetector based on single crystalline MoO$_3$ nanosheets. *Journal of Materials Chemistry C*, *3*(28), 7469–7475. https://doi.org/10.1039/C5TC00850F
48. Zheng, B., Wang, Z., Chen, Y., Zhang, W., & Li, X. (2018). Centimeter-sized 2D α-MoO3 single crystal: Growth, Raman anisotropy, and optoelectronic properties. *2D Materials*, *5*(4), 045011. https://doi.org/10.1088/2053-1583/aad2ba
49. Jones, H. A., R, I. L., & Mackay, G. M. J. (1927). The Rates of Evaporation and the Vapor Pressures of Tungsten, Molybdenum, Platinum, Nickel, Iron, Copper and Silver. *Physical Review*, *30*(2), 201–214. https://doi.org/10.1103/PhysRev.30.201
50. Bica de Moraes, M. A., Trasferetti, B. C., Rouxinol, F. P., Landers, R., Durrant, S. F., Scarmínio, J., & Urbano, A. (2004). Molybdenum Oxide Thin Films Obtained by the Hot-Filament Metal Oxide Deposition Technique. *Chemistry of Materials*, *16*(3), 513–520. https://doi.org/10.1021/cm034551a
51. Siciliano, T., Tepore, A., Filippo, E., Micocci, G., & Tepore, M. (2009). Characteristics of molybdenum trioxide nanobelts prepared by thermal evaporation technique. *Materials*

*Chemistry and Physics*, *114*(2), 687–691. https://doi.org/10.1016/j.matchemphys.2008.10.018
52. Gong, Y., Zhao, Y., Zhou, Z., Li, D., Mao, H., Bao, Q., Zhang, Y., & Wang, G. P. (2022). Polarized Raman Scattering of In-Plane Anisotropic Phonon Modes in α-MoO$_3$. *Advanced Optical Materials*, *10*(10), 2200038. https://doi.org/10.1002/adom.202200038
53. Wen, M., Chen, X., Zheng, Z., Deng, S., Li, Z., Wang, W., & Chen, H. (2021). In-Plane Anisotropic Raman Spectroscopy of van der Waals α-MoO$_3$. *The Journal of Physical Chemistry C*, *125*(1), 765–773. https://doi.org/10.1021/acs.jpcc.0c09178
54. Jain, A., Ong, S. P., Hautier, G., Chen, W., Richards, W. D., Dacek, S., Cholia, S., Gunter, D., Skinner, D., Ceder, G., & Persson, K. A. (2013). Commentary: The Materials Project: A materials genome approach to accelerating materials innovation. *APL Materials*, *1*(1), 011002. https://doi.org/10.1063/1.4812323
55. Horton, M. K., Huck, P., Yang, R. X., Munro, J. M., Dwaraknath, S., Ganose, A. M., Kingsbury, R. S., Wen, M., Shen, J. X., Mathis, T. S., Kaplan, A. D., Berket, K., Riebesell, J., George, J., Rosen, A. S., Spotte-Smith, E. W. C., McDermott, M. J., Cohen, O. A., Dunn, A., … Persson, K. A. (2025). Accelerated data-driven materials science with the Materials Project. *Nature Materials*, 1–11. https://doi.org/10.1038/s41563-025-02272-0
56. Andersson, G., & Magneli, A. (1950). On the crystal structure of molybdenum trioxide. *Acta Chemica Scandinavica*, *4*, 793–797. https://doi.org/10.3891/acta.chem.scand.04-0793
57. Sitepu, H. (2009). Texture and structural refinement using neutron diffraction data from molybdite (MoO$_3$ ) and calcite (CaCO$_3$ ) powders and a Ni-rich Ni$_{50.7}$Ti$_{49.30}$ alloy. *Powder Diffraction*, *24*(4), 315–326. https://doi.org/10.1154/1.3257906
58. Dieterle, M., Weinberg, G., & Mestl, G. (2002). Raman spectroscopy of molybdenum oxides. *Physical Chemistry Chemical Physics*, *4*(5), 812–821. https://doi.org/10.1039/b107012f
59. Firment, L. E., & Ferretti, A. (1983). Stoichiometric and oxygen deficient MoO3(010) surfaces. *Surface Science*, *129*(1), 155–176. https://doi.org/10.1016/0039-6028(83)90100-0
60. Markov, I. V. (2017). *Crystal growth for beginners: Fundamentals of nucleation, crystal growth, and epitaxy* (3rd edition). World Scientific. https://doi.org/10.1142/10127
61. Kita, N. T., Ushikubo, T., Fu, B., & Valley, J. W. (2009). High precision SIMS oxygen isotope analysis and the effect of sample topography. *Chemical Geology*, *264*(1), 43–57. https://doi.org/10.1016/j.chemgeo.2009.02.012
62. Tang, G.-Q., Li, X.-H., Li, Q.-L., Liu, Y., Ling, X.-X., & Yin, Q.-Z. (2015). Deciphering the physical mechanism of the topography effect for oxygen isotope measurements using a Cameca IMS-1280 SIMS. *Journal of Analytical Atomic Spectrometry*, *30*(4), 950–956. https://doi.org/10.1039/C4JA00458B